%% file: conference.tex
\documentclass[conference]{IEEEtran}
\IEEEoverridecommandlockouts
\usepackage{cite}
\PassOptionsToPackage{hyphens}{url}\usepackage{hyperref}
\usepackage{amsmath,amssymb,amsfonts}
\usepackage{algorithmic}
\usepackage{graphicx}
\usepackage{textcomp}
\usepackage{xcolor}
\usepackage{float}
\usepackage{svg}
\usepackage{siunitx}
\emergencystretch=\maxdimen
\usepackage{cuted}
\usepackage{multirow}
\usepackage{longtable}
\usepackage[export]{adjustbox}
\usepackage{graphicx, subcaption, caption}
\usepackage{titlesec}
\titlespacing*{\section}
{0pt}{6pt}{6pt}
\titlespacing*{\subsection}
{0pt}{6pt}{2pt}
\usepackage{lipsum}
\usepackage{textcomp}
\usepackage{xcolor}
\usepackage{siunitx}
\usepackage[bottom]{footmisc}
\usepackage[utf8]{inputenc}
\usepackage[T1]{fontenc}
\usepackage{booktabs, makecell, ltablex, threeparttablex}

\def\BibTeX{{\rm B\kern-.05em{\sc i\kern-.025em b}\kern-.08em
    T\kern-.1667em\lower.7ex\hbox{E}\kern-.125emX}}
\begin{document}

\title{Novel Simplified Model for Planar Capacitors up to 1 GHz

 \thanks{This work was partially funded by CAPES, INCT-NAMITEC (CNPq No. 406193/2022-3), and MANNA-BR/@manna Team, with support from CNPq, Fundação Araucária, Softex, and MCTI.}% <-this % stops a space
}

% AUTORES (menor) 
\author{
\IEEEauthorblockN{
Reinan Lima\IEEEauthorrefmark{1}, 
Graziella Bedenik\IEEEauthorrefmark{2}, 
Farshad Yazdani\IEEEauthorrefmark{1},  
Matthew Robertson\IEEEauthorrefmark{2}, 
and Elyson Ádan Nunes Carvalho\IEEEauthorrefmark{1}}

\IEEEauthorblockA{\IEEEauthorrefmark{1}Universidade Federal de Sergipe, São Cristóvão, Brazil \\
reinanbom@gmail.com, \{farshad, ecarvalho\}@academico.ufs.br }

\IEEEauthorblockA{\IEEEauthorrefmark{2}Queen's University, Kingston, Canada \\
\{graziella.bedenik, m.robertson\}@queensu.ca}
}

\maketitle

\begin{abstract}
The study and modelling of planar capacitors remain highly relevant due to their broad applications. However, their accurate modelling presents significant challenges due to intrinsic high-frequency effects, which include not only parasitic inductance but also ohmic losses in the copper and dielectric losses in the substrate. Therefore, this work proposes developing a new simplified planar capacitor model consisting of two identical rectangular plates, which incorporates the effects of both the substrate and the medium between the conductor traces on the total capacitance. Moreover, it includes the impact of parasitic inductance and losses associated with the copper and substrate. This study demonstrated the importance of modelling both the dielectric substrate and the inter-conductor medium, with results showing significant improvements in the proposed model's accuracy. Experimental validation through comparison with an already established model in the literature and physical planar capacitors revealed good agreement for most cases studied.
\end{abstract}

\begin{IEEEkeywords}
Planar Capacitor, Modelling, High-frequency
\end{IEEEkeywords}

\input{text/intro}
\input{text/model}

\input{text/experiments}

\input{text/results}
\input{text/conclusion}

\section*{Acknowledgement}

We thank Diancheng Li for assisting with sensor manufacturing.

\bibliographystyle{IEEEtran}
\bibliography{bibliography}

\end{document}

%% file: text/intro.tex
\section{Introduction}
\label{sec:intro}

The study and modeling of planar capacitors remain highly relevant due to their widespread use across various applications, particularly in microelectronics and sensing technologies \cite{zhao2009field,wang1999modeling,hirsch2019capacitive,abdelraheem2021iot,koloskov2024modeling,boonkirdram2015novel}. Their low cost, fast response, and flexible electrode configurations make them especially attractive for sensor development \cite{shi1991capacitance,huang1989tomographic}. At the same time, unintentional parasitic capacitances—often introduced by PCB traces that behave like planar capacitors—can significantly influence the performance of electronic circuits and measurement systems \cite{pajovic2007analysis}. As a result, a deeper understanding and accurate modelling of planar capacitors is crucial not only for intentional sensor design but also for minimizing unwanted parasitic effects.

\begin{figure}[t]
  \centering
  \includegraphics[width=0.47\textwidth]{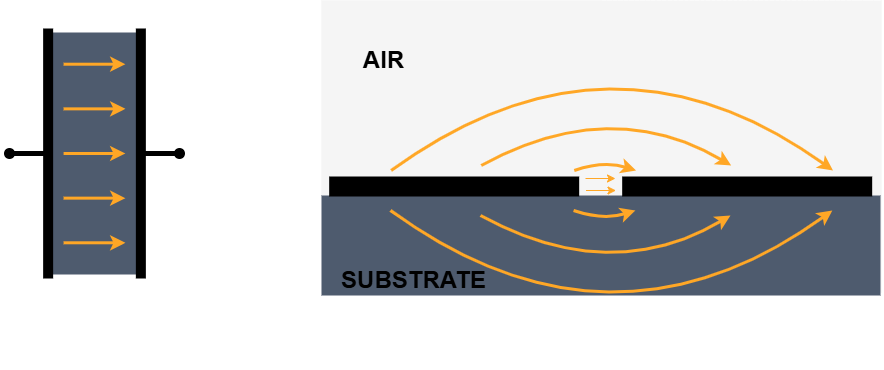}
  \caption{On the left, a parallel-plate capacitor with uniform electric field, neglecting edge effects. On the right is a planar capacitor with coplanar electrodes on a substrate, where the electric field exhibits curvature and non-uniformity, highlighting the challenges in electrostatic modelling of this configuration.}
  \label{fig:caps}
\end{figure}

Despite being a well-established topic, planar capacitor modelling still faces significant challenges, particularly when accounting for measurement phenomena at frequencies in the range of tens to hundreds of MHz. Many existing models oversimplify the physical behaviour by neglecting subtle yet critical effects such as parasitic inductance, ohmic losses in conductive traces, and dielectric losses in the substrate material \cite{tria2016planar,isidoro2017scalable}. These factors limit the applicability of the classical parallel-plate capacitor equation ($C = \epsilon_0 \frac{A}{d}$), which assumes uniform field distributions that do not hold in planar geometries \cite{wei2017modeling}, as shown in Fig.~\ref{fig:caps}. To address these limitations, three main strategies have emerged in the literature: (1) empirical corrections to traditional formulas through fitting parameters and adjustment factors \cite{allstot2003parasitic,wintle1989capacitance,kuester1988explicit}; (2) numerical simulations (e.g., finite element method or finite difference time domain approaches) that solve Maxwell's equations directly for complex geometries, and (3) the development of new formulations that explicitly incorporate high-frequency and geometric effects. Although the latter is more complex, it offers improved accuracy and physical insight, making it particularly valuable for modern high-frequency applications. Numerical simulations, while accurate, often obscure fundamental parametric relationships within localized results, making it difficult to extract general trends—a limitation that analytical models overcome.

% However, accurate modelling of planar capacitors presents significant challenges due to intrinsic high-frequency effects, which include not only parasitic inductance but also ohmic losses in the copper and dielectric losses in the substrate \cite{tria2016planar,isidoro2017scalable}. These phenomena render the conventional parallel-plate formula ($C=\epsilon_0 \frac{A}{d}$) inadequate, since planar capacitors exhibit substantially different field distributions \cite{wei2017modeling}, as illustrated in Fig.~\ref{fig:caps}. Consequently, two main approaches emerge in the literature: 1) employing empirical methods that adjust existing formulas through correction terms and constants \cite{allstot2003parasitic,wintle1989capacitance,kuester1988explicit}, and 2) developing completely new formulations that explicitly incorporate high-frequency effects. Although more complex, this second approach provides greater accuracy and physical insight.

\begin{figure}[t]
  \centering
  \includegraphics[width=0.3\textwidth]{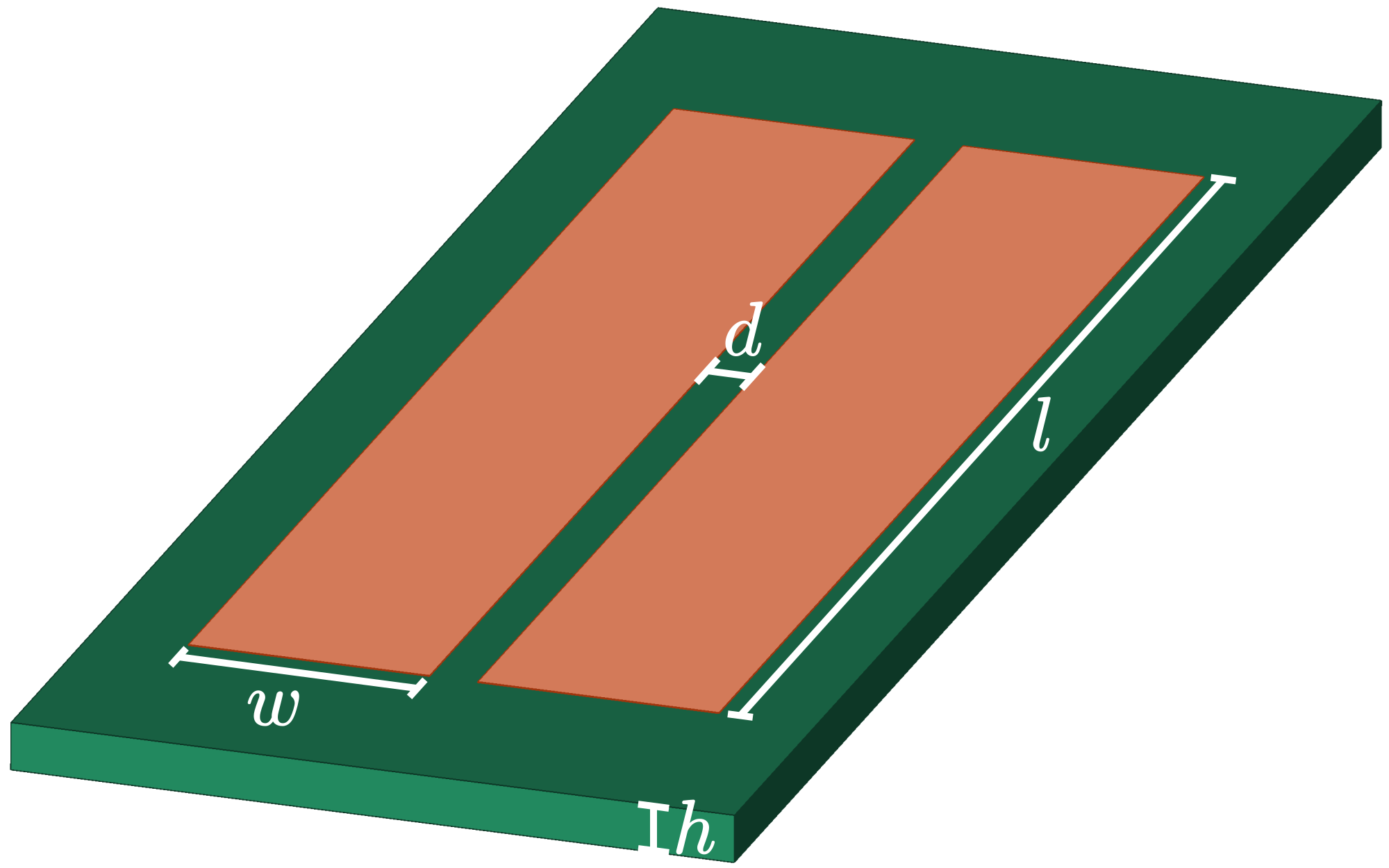}
  \caption{Two copper plates on a substrate, with equal widths ($W$), trace spacing ($d$), length ($l$), and substrate height ($h$).}
  \label{fig:3Dcaps}
\end{figure}

A review of existing works reveals varying levels of depth in addressing these effects. In \cite{teodorescu2024constraints}, Teodorescu only superficially mentions parasitic inductance and losses, while Vukadinovic \textit{et al.} \cite{vukadinovic2009modelling} and Deleniv \cite{deleniv1999question} primarily focus on capacitance formulations without fully considering high-frequency regimes. Vendik \textit{et al.}  \cite{vendik1999modeling} provides a more comprehensive contribution. Their model explicitly accounts for both the surrounding medium and the dielectric substrate through the expression
\begin{equation}
\label{eq:vendik}
C = \frac{2 \varepsilon_0 l}{\pi} \ln\left( \frac{4(d + 2W)}{d} \right) + \frac{\varepsilon_0 l (\varepsilon_r - 1)}{\pi} \ln\left( \frac{16h}{\pi d} \right),
\end{equation}
where $C$ is capacitance, $\varepsilon_r$ is the substrate's electric permeability, $\varepsilon_0$ is the vacuum electric permeability, and the other dimensional parameters are illustrated in Fig. \ref{fig:3Dcaps}. Although comprehensive, \eqref{eq:vendik} still fails to incorporate the high-frequency effects that become relevant in the MHz range.

%Deleniv at al. \cite{deleniv1999question} apresenta fórmula e assume eletrodos semi-infinitos, ou seja, muito longos comparados à separação entre eles ou à espessura das camadas, que é expresso como:

%\begin{equation}\small
% C = \frac{\epsilon l}{\pi} \left[ \ln\left(\dfrac{16h}{\pi d} \right) \right]
%\end{equation}

This work introduces a simplified yet comprehensive analytical model for planar capacitors, developed to support the simulation and analysis of measurement systems operating below 1 GHz. The primary contribution of this paper is the development of an analytical model that incorporates the effects of both the substrate and the medium between conductor traces on the total capacitance of planar capacitors—factors often overlooked or treated separately in previous work. We consolidate and extend several equation fragments from the literature into a unified analytical expression capable of estimating the capacitor’s impedance. While numerical simulations can yield higher accuracy, the proposed analytical model offers key advantages: (i) it reveals explicit relationships between physical parameters, enabling identification of dominant effects across frequency ranges; (ii) it provides a computationally efficient starting point for early-stage design, particularly when multi-physics simulations are too resource-intensive, and (iii) it enables predicting the influence of geometry and dielectric properties on the component, which is especially beneficial for sensing applications.

%% file: text/model.tex
\section{Model Development}
\label{sec:methods}

The conventional capacitance formula, valid for ideal parallel-plate capacitors (Fig.~\ref{fig:caps}, left), proves inadequate for planar capacitors (Fig.~\ref{fig:caps}, right), particularly at high frequencies. The electric field lines exhibit curvature in the planar configuration, demonstrating a non-uniform distribution that invalidates the classical concept of well-defined area \(A\) and separation \(d\). Furthermore, dynamic effects, including parasitic inductance and frequency-dependent material resistance, become significant in the system, requiring models incorporating geometric corrections and coupled electromagnetic behaviour for physically consistent characterization.

\begin{figure}[t]
  \centering
  \includegraphics[width=0.3\textwidth]{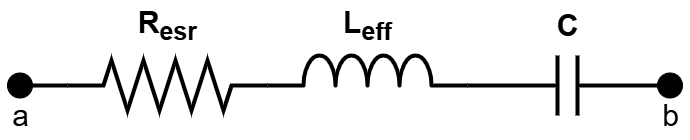}
  \caption{RLC model of a planar capacitor with electrodes on a substrate between terminals a and b.}
  \label{fig:modelo}
\end{figure}

The complete system exhibits combined behaviour from three fundamental elements. Each plate contributes a series parasitic inductance ($L_{eff}$) due to the closed current path. At the same time, we now have a series resistance ($R_{esr}$) due to the copper resistance ($R_{cu}$) which becomes significant at high frequencies due to the skin effect, and the loss through the substrate which can be modelled as resistance ($R_{sub}$). Despite these parasitic effects, the capacitance ($C$) remains the dominant system component until a certain point. This interaction results in an equivalent series RLC circuit, as established in specialized literature \cite{neu2003designing,winslow2000component}, whose schematic representation is shown in Fig. \ref{fig:modelo}.

Subsections \ref{subsec:subcaps}, \ref{subsec:induc}, and \ref{subsec:res} separately analyze the components of each sub-model based on the capacitor geometry illustrated in Fig.~\ref{fig:3Dcaps}.

\subsection{Capacitance}
\label{subsec:subcaps}
\indent

Since we are analyzing a planar capacitor, capacitance is the principal element in our modelling. As illustrated in Fig.~\ref{fig:caps}, unlike parallel-plate capacitors, planar capacitors exhibit electric field lines distributed through both air and substrate. Furthermore, although the conductor trace thickness is relatively small (typically 35 $\mu$m), it affects the electric field distribution, enabling field lines to form between traces through the copper height, which contributes to the system's total capacitance. For this reason, we propose a model incorporating the combined effects of these three regions: air, substrate, and the medium between traces.

The capacitance $C$ was divided into two components: $C_1$, representing the substrate capacitance,
\begin{equation}\small
\label{Csub}
C_1 = \frac{\varepsilon_{sub} \varepsilon_0 l}{\pi} \ln\left(\dfrac{2W}{d} + 1\right),
\end{equation}
and $C_2$, accounting for the air region and inter-conductor medium,
\begin{equation}\small
\label{Cair}
C_2 = \frac{\varepsilon_{r} \varepsilon_0 l}{\pi-\varphi} \ln\left(\dfrac{2W}{d} + 1\right),
\end{equation}
where $\varepsilon_{sub}$ is the substrate permittivity, $\varepsilon_{0}$ the vacuum permittivity, and $\varepsilon_{r}$ the surrounding medium's permittivity ($\varepsilon_r = 1$ for air). The total capacitance combines both contributions:
\begin{equation}\small
\label{meuC}
C = C_1 + C_2 = \varepsilon_0 l \ln\left(\dfrac{2W}{d} + 1\right) \left[\frac{\varepsilon_{sub}}{\pi}+\frac{\varepsilon_r}{\pi -\varphi}\right],
\end{equation}
\begin{equation}\small
\label{meuc2}
\varphi = \tan^{-1}\left(\frac{2t}{W + d}\right),
\end{equation}
where $t$ represents the thickness of the copper.

\subsection{Inductance}
\label{subsec:induc}

Current flow through both conducting plates generates an associated magnetic field, introducing parasitic inductance effects in the system. While this phenomenon is often negligible at low frequencies, it becomes significant at high frequencies \cite{griffiths2023introduction}, necessitating its incorporation in the proposed model.

To approximate this parasitic inductance, we adopt formulations available in the literature \cite{bueno1995new,paul2011inductance}. The individual inductance $L$ of each plate is calculated by
\begin{equation}\small
\label{eq5}
L = \frac{\mu_0 l}{2\pi}\left[2\ln\left(\frac{2l}{W}\right) - 2\right],
\end{equation}
where $\mu_0$ is the vacuum magnetic permeability.

Considering that the currents in the plates form a closed loop, the combined effect results in a total effective inductance given by
\begin{equation}\small
\label{eq6}
L_{eff} = 2L.
\end{equation}

This inductance doubling reflects the series contribution of both conductors in the current path.

\subsection{Resistance}
\label{subsec:res}

The copper plate resistance ($R_{cu}$) exhibits strong frequency dependence due to the skin effect \cite{ramo1994fields}, a phenomenon particularly relevant in high-frequency systems like our model. As frequency increases, the current density becomes progressively concentrated near the conductor surface, reducing the effective conduction area \cite{griffiths2023introduction}. This behaviour is quantified by the skin depth $\delta$, representing the penetration distance where current density decays to $\frac{1}{e}$ (approximately 37\%) of its surface value, according to
\begin{equation}\label{skin_depth}\small
\delta = \frac{1}{\sqrt{\pi f \mu_0 \sigma}},
\end{equation}
where $f$ is the operating frequency and $\sigma$ is the copper conductivity.

For our high-frequency planar capacitor, the effective copper resistance is given by \cite{griffiths2023introduction}
\begin{equation}\label{eq9}\small
R_{cu} = \frac{l}{\delta W \sigma}.
\end{equation}

The dielectric substrate losses are modelled using the formula presented and discussed by Deleniv \cite{deleniv2009modeling}, which relates the substrate resistance to the material's dissipation factor,
\begin{equation}\label{eq:sub_loss}\small
R_{sub} = \frac{\tan\delta}{2\pi f C}
\end{equation}
where $\tan\delta$ represents the substrate's dissipation factor, with a typical value of 0.02 for FR-4 materials in the RF range. Since the system contains two identical conductive plates, the total copper resistance is naturally doubled, resulting in $2R_{cu}$ for the complete set. Consequently, the total equivalent series resistance ($R_{esr}$) that incorporates both loss mechanisms is then given by
\begin{equation}\label{eq:total_esr}\small
R_{esr} = 2R_{cu} + R_{sub}.
\end{equation}

\subsection{Total Impedance}
\indent 

With the three main model components already determined, ($R_{esr}$, $L_{eff}$, and $C$), we can now define the complete impedance of the planar capacitor. This total impedance incorporates both the system's resistive losses and parasitic effects, expressed by the vector sum of individual contributions from each element, being described as
\begin{equation}\label{eq12}\small
Z_{total} = R_{esr} + j\omega L_{eff} + \frac{1}{j\omega C}.
\end{equation}

This formulation enables a comprehensive analysis of the capacitor's behaviour, being particularly relevant for determining the system's resonance frequency. At this critical point, inductive and capacitive reactances mutually cancel each other. Identifying this parameter is crucial as it establishes the boundary between two distinct operational regions: below resonance, where the desired capacitive behaviour predominates, and above resonance, where parasitic inductive effects become dominant.

% \begin{equation}\small
% \begin{aligned}
% Z_{total} =\ & 2\left( \frac{l}{\delta W \sigma} \right) 
% + \frac{\tan\delta}{2\pi f \varepsilon_0 l \ln\left( \dfrac{2W}{d} + 1 \right) \left[ \dfrac{\varepsilon_{sub}}{\pi} + \dfrac{\varepsilon_r}{\pi - \tan^{-1}\left( \dfrac{2t}{W + d} \right)} \right]} \\
% & +\ j\omega \left( \frac{\mu_0 l}{\pi} \left[\ln\left( \frac{2l}{W} \right) - 1 \right] \right) \\
% & +\ \frac{1}{j\omega \varepsilon_0 l \ln\left( \dfrac{2W}{d} + 1 \right) \left[ \dfrac{\varepsilon_{sub}}{\pi} + \dfrac{\varepsilon_r}{\pi - \tan^{-1}\left( \dfrac{2t}{W + d} \right)} \right]}
% \end{aligned}
% \end{equation}

%% file: text/experiments.tex
\section{Experimental Methodology}
\label{sec:experiments}

The experimental validation focuses on the proposed series RLC model defined in \eqref{eq12}, along with its associated capacitance expressions \eqref{meuC} and \eqref{meuc2}. To assess the model’s accuracy, we conducted systematic comparisons against both the established model by Vendik \textit{et al.} \cite{vendik1999modeling} and experimental measurements from real planar capacitors. As this study emphasizes analytical modelling, electromagnetic simulations were not included; instead, the validation relies directly on physical component measurements. The full methodology is described in subsections \ref{subsec:simulations} and \ref{subsec:real_caps}.
\subsection{Analytical Models Simulation}
\label{subsec:simulations}

Two different rounds of analytical model simulations were implemented in GNU Octave to enable a comparative approach between the two impedance models across a frequency range from 100 kHz to 1 GHz. Both models share the same formulations for parasitic inductance calculation, described in \eqref{eq6}, and series resistance, described in \eqref{eq:total_esr}. The difference lies in the capacitance calculation, where one model employs the formulation we propose in section \ref{subsec:subcaps} as \eqref{meuC} and \eqref{meuc2}, and the other uses Vendik et al.'s formula, presented in \eqref{eq:vendik}.

For this analysis, we used the real part of the reflection coefficient $S_{11}$, calculated by
\begin{equation}
S_{11_{real}} = \Re\left[\frac{Z - Z_0}{Z + Z_0}\right],
\end{equation}
where $Z$ is the total system impedance and $Z_0 =$ 50 $\Omega$.
% W = 100mm, l = 800mm, d = 10mm

Initially, we kept the geometric parameters shown in Fig.~\ref{fig:3Dcaps} identical while intentionally varying the substrate permittivity. This approach enables the evaluation of the sensitivity of the proposed models to the dielectric material properties—an important aspect, as both formulations incorporate the contribution of field lines passing through the substrate. For the first simulation, we adopted the typical relative permittivity of FR-4 ($\varepsilon_r = 4.3$), a material widely used in high-frequency applications. In contrast, the second simulation employed Teflon ($\varepsilon_r = 2.1$), selected for its distinct dielectric properties, thereby enabling a comparative analysis of the results' dependence on the substrate dielectric constant.

In the second round of simulations, three planar capacitor geometric configurations were selected for systematic evaluation of the proposed model. These configurations, presented in Table~\ref{tab:capacitores}, were carefully chosen to cover a wide range of physical parameters. Capacitor 1, with a surface area of \SI{1600}{\milli\meter\squared}, represents the largest configuration. Capacitor 2 features a reduced area of \SI{320}{\milli\meter\squared}, enabling evaluation of the model's performance in elongated, narrow geometries. Capacitor 3 provides an intermediate case with \SI{600}{\milli\meter\squared} area for balanced comparative analysis. This systematic variation enables an assessment of the model's robustness against changes in geometric parameters.

\begin{table}[h]
\caption{Dimensional Parameters (mm) for the Design of Simulated Capacitors}
\centering
\begin{tabular}{c c c c c}
\hline
\textbf{Capacitor Design} & \textbf{$W$} & \textbf{$l$} & \textbf{$d$} & \textbf{$h$} \\
\hline
1 & 20 & 80 & 1 & 2 \\
2 & 4  & 80 & 2 & 2 \\
3 & 10 & 60 & 2 & 2 \\
\hline
\end{tabular}

\label{tab:capacitores}
\end{table}

\subsection{Comparison with Real Capacitors}
\label{subsec:real_caps}

To enable comparison with real components and experimental validation of the proposed model, we fabricated physical prototypes of planar capacitors, which are shown in \ref{fig:3Dcaps}. Their design was implemented according to the configurations in Table \ref{tab:capacitores} to ensure comparability between theoretical and experimental results. We used 2 mm FR4 substrate with a $35\,\mu\text{m}$ copper layer. The fabrication method was milling through a CNC machine (ProtoMat S64, LPKF Laser \& Electronics, Germany). 

\begin{figure}[b]
  \centering
  \includegraphics[width=0.33\textwidth]{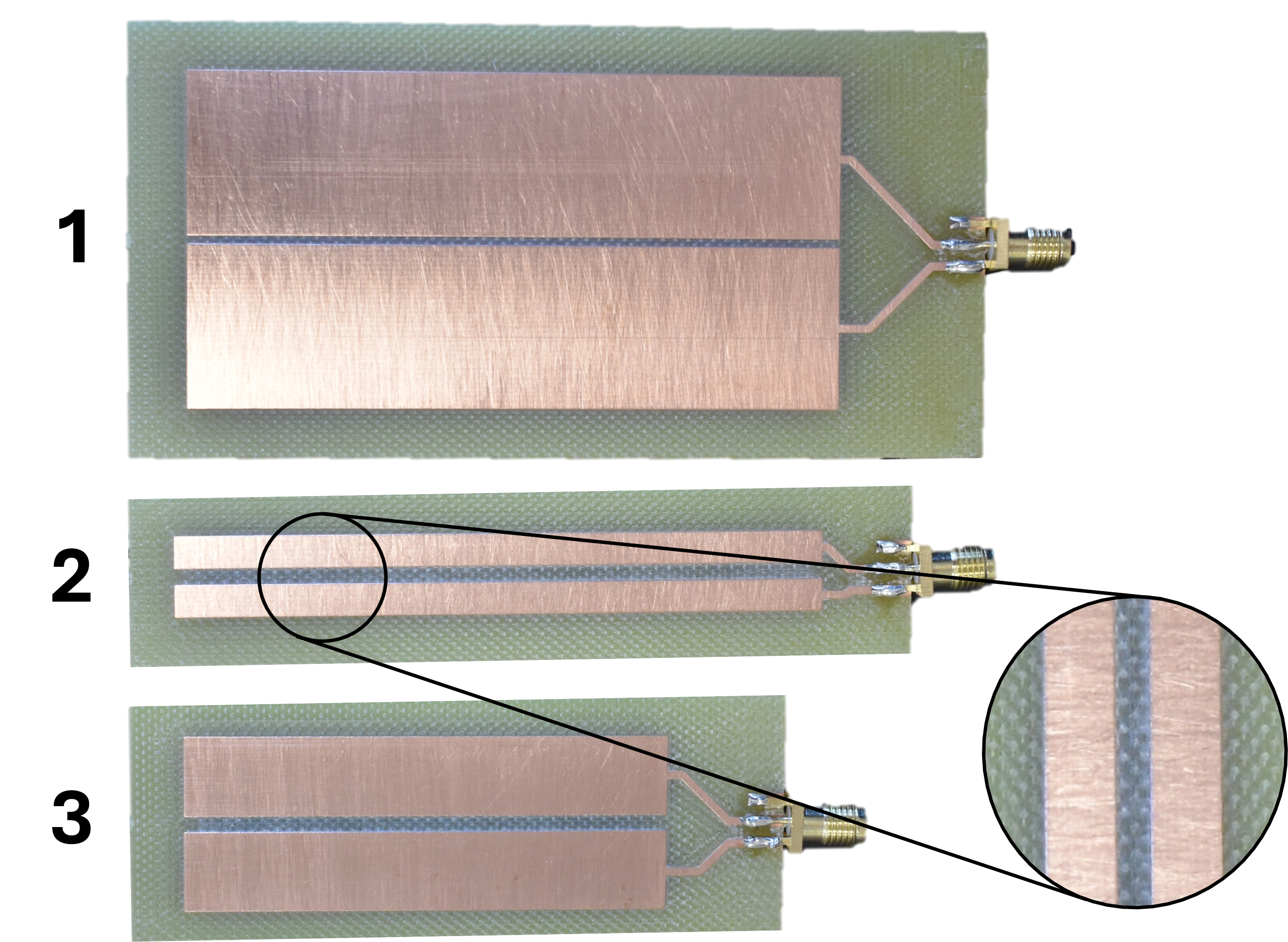}
  \caption{Fabricated planar capacitor designs, showcasing the precise geometry and clean and polished surface finish.}
  \label{fig:caps_reais}
\end{figure}

The capacitors underwent a rigorous cleaning process, which included mechanical removal of the oxidized copper layer using steel wool and chemical cleaning with isopropyl alcohol. The electrical integrity of each prototype was verified through continuity measurements with a digital multimeter and visual inspection under an optical microscope to detect potential short circuits or surface imperfections. Finally, female SMA connectors were soldered on each board.

Characterization of $S$ parameters was performed using a vector network analyzer (NanoVNA-H, v1.2.27) calibrated for the frequency range of interest (matching the simulation range from subsection \ref{subsec:simulations}), collecting 6401 data points. Both real and imaginary $S_{11}$ parameters were acquired using the NanoVNA-App by OneOfEleven (v1.1.208), interfaced with the device.

The experimental impedance ($Z$) was determined from the fundamental relation
\begin{equation}
Z = Z_0\frac{1 - \Gamma}{1 + \Gamma},
\end{equation}
where the complex reflection coefficient $\Gamma = S_{11_{\text{real}}} + jS_{11_{\text{im}}}$ incorporates both measured components of the $S_{11}$ parameter. This conversion enables direct comparison with impedances calculated by the analytical models.

%% file: text/results.tex
\section{Results and Discussion} 
\label{sec:results}

Following the experimental methods described in section \ref{sec:experiments}, the results analysis focused on quantitative model validation. For that, two main aspects were involved. The first is calculating the mean squared error between measured and modelled impedances,
\begin{equation}
\text{MSE(\%)} = \left[\frac{1}{N} \sum_{i=1}^{N} \left(   Z_{1,i} - Z_{2,i}  \right)^2\right]\cdot100,
\end{equation}
where $Z_1$ and $Z_2$ are impedances and $N$ is the number of samples in the vector, which objectively quantifies the global discrepancy between experimental data and theoretical predictions. Secondly, a comparative analysis of resonance frequencies was performed, as this critical parameter marks the transition between the device's capacitive and inductive operation regimes.

% This section is organized into two parts: (\ref{res}) the obtained results, derived from the experiments conducted in section~\ref{sec:experiments}; and (\ref{dis}) the conclusions, featuring a critical analysis of the results and their implications for the proposed model.

\begin{figure}[b]
  \centering
  \includegraphics[width=0.5\textwidth]{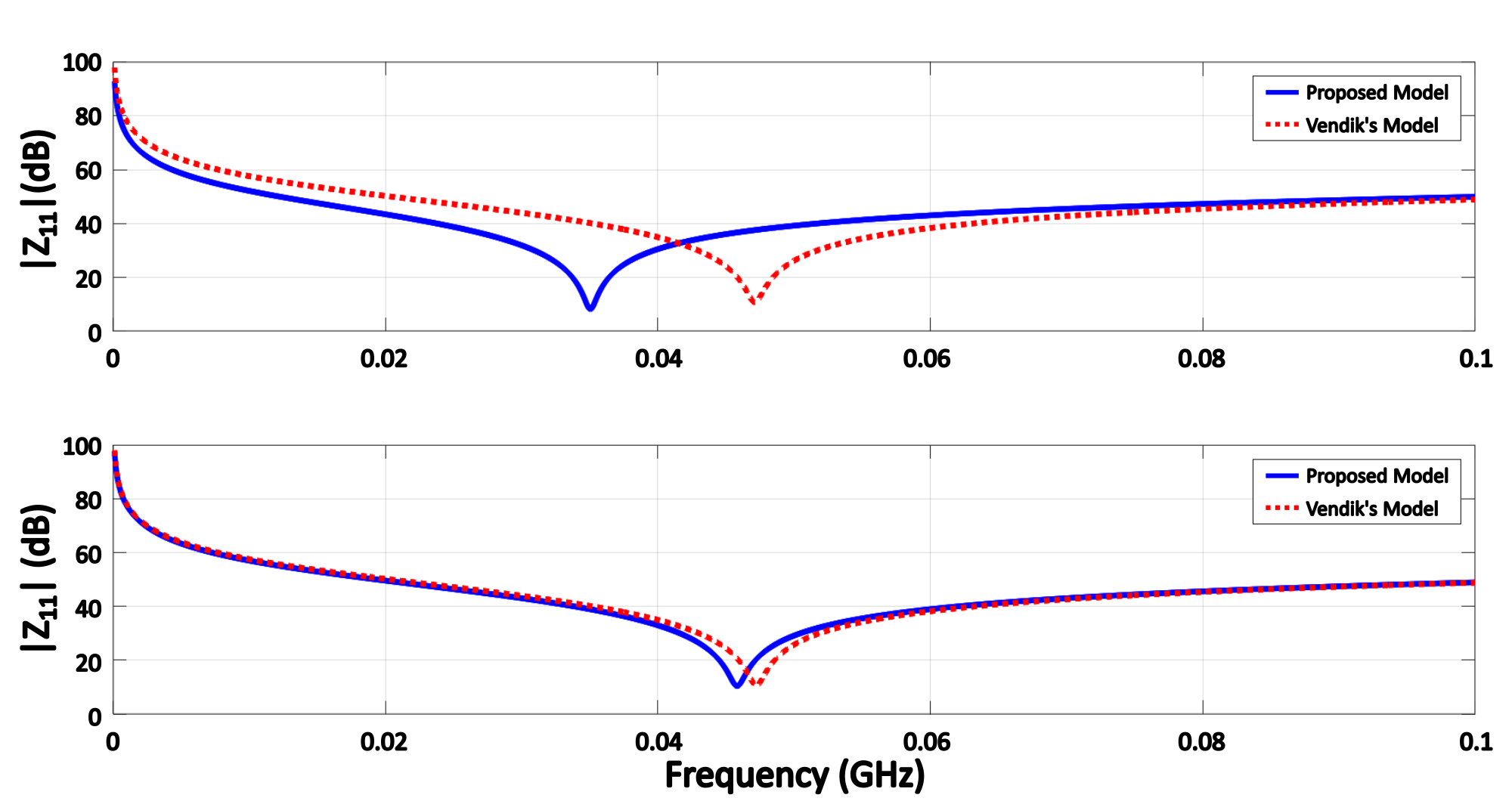}
  \caption{Impedance's comparison curves of the models, with the same dimensions, but varying the electrical permittivity of the substrate. FR-4 on top and Teflon on bottom.}
  \label{fig:impedancia_dif_sub}
\end{figure}

The results illustrated in Fig.~\ref{fig:impedancia_dif_sub} show distinct behaviours between the two analytical models regarding substrate permittivity variation. The proposed model exhibits a noticeable shift in resonance frequency when replacing the FR-4 substrate ($\varepsilon_r = 4.3$) with Teflon ($\varepsilon_r = 2.1$). This displacement agrees with the expected reduction in effective capacitance, since capacitance is directly proportional to the dielectric medium's permittivity. The proposed model's sensitivity to material properties demonstrates its ability to accurately capture the substrate's influence on capacitor behaviour, which is a key design aspect, as selecting substrates with different dielectric properties would yield significantly different responses if this parameter weren't adequately considered. In contrast, Vendik's model shows practically unchanged response to the same material variation, demonstrating its limitation in adequately incorporating substrate effects. This behavioural difference highlights the advantage of the proposed model, which more completely considers field lines traversing both the substrate and the space between copper traces, thereby capturing the actual influence of dielectric properties on component performance. The proposed model's ability to appropriately respond to these variations demonstrates its superior performance and utility for designs where substrate material selection is a relevant design parameter, which reinforces our initial argument that existing models often neglect subtle, but essential effects, explored in section \ref{sec:intro}.

\begin{figure}[b]
  \centering
  \includegraphics[width=0.5\textwidth]{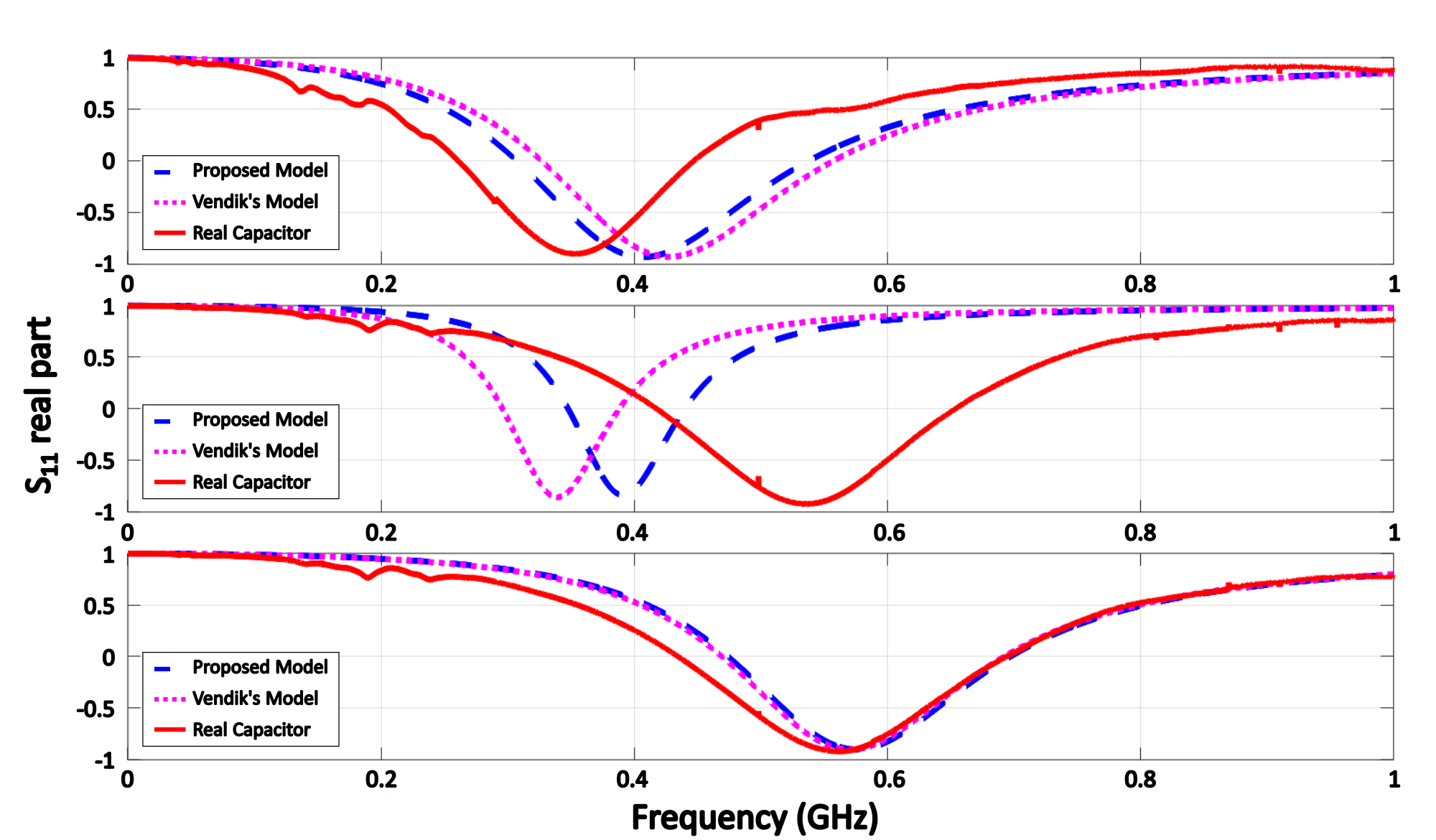}
  \caption{Comparison of the real part of the reflection coefficient $S_{11}$ between models and physical capacitor measurements. From top to bottom: capacitor 1, capacitor 2, and capacitor 3.}
  \label{fig:imagem_S11}
\end{figure}

\begin{figure}[t]
  \centering
  \includegraphics[width=0.5\textwidth]{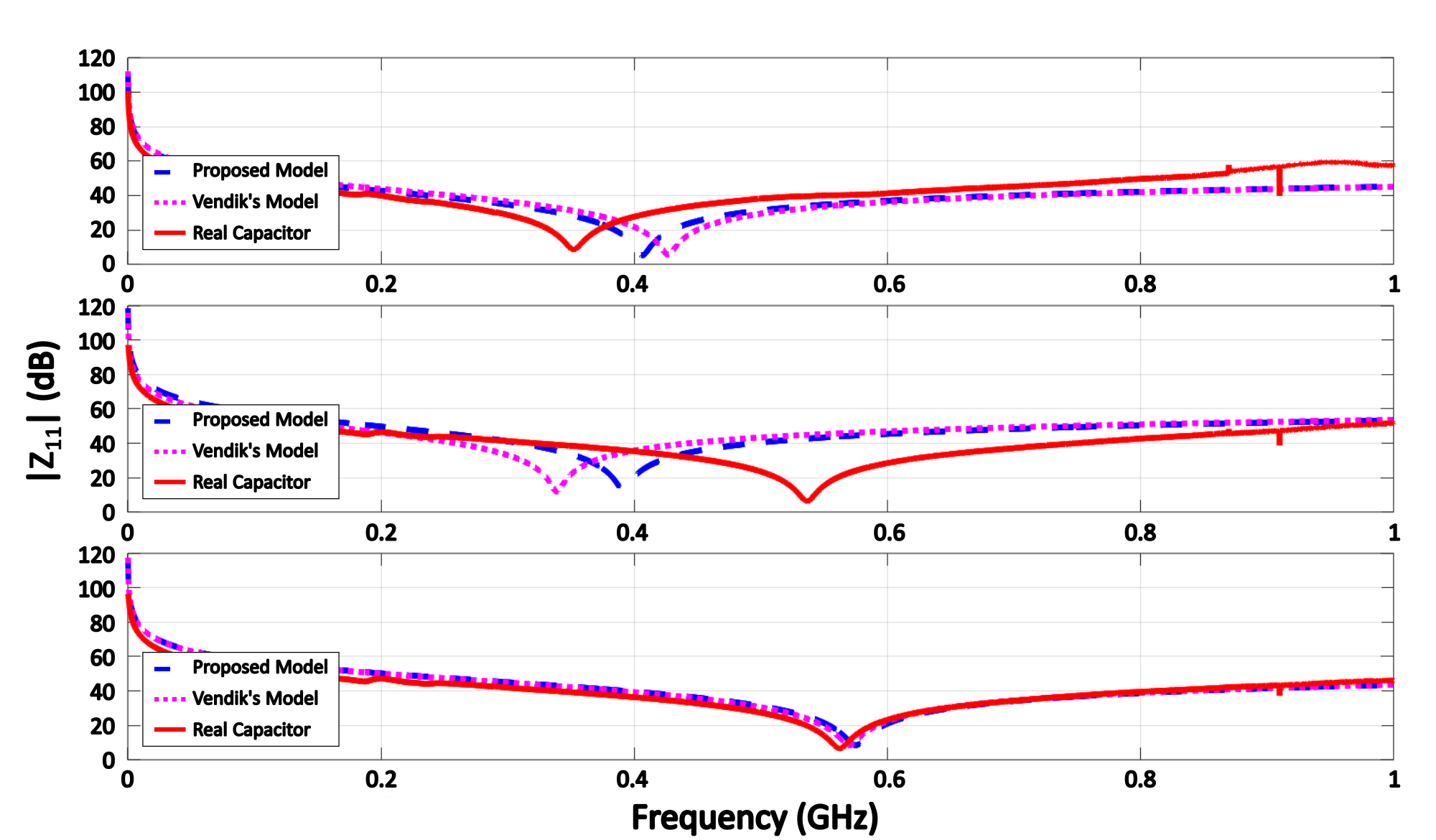}
  \caption{Comparison of impedance magnitude curves between models and physical capacitor measurements. From top to bottom: capacitor 1, capacitor 2, and
capacitor 3.}
  \label{fig:imagem_Z11}
\end{figure}

Figures~\ref{fig:imagem_S11} and~\ref{fig:imagem_Z11}, depicting respectively the real part of $S_{11}$ and the reflection impedance magnitude ($|Z_{11}|$) over the frequency range, present the results obtained for each evaluated capacitor, whose dimensions are listed in Table~\ref{tab:capacitores}. The mean squared errors between the models and the physical capacitors are reported in Table \ref{tab:erro}. At the same time, the measured and simulated resonance frequencies are shown in Table \ref{tab:freq}. Using these metrics for a comparative analysis reveals important relationships between different geometries and model performance. 

Beginning with Capacitor 1, good agreement was observed between theory and experiment. In this case, the proposed model demonstrated better performance than Vendik's, with a mean squared error of 1.060\% against 1.323\%. The experimental resonance frequency of 352.4 MHz was predicted with an error of 15.5\% (407.1 MHz) by the proposed model and 21.3\% (427.4 MHz) by Vendik's model, indicating a curve shift. 

For Capacitor 2, both models significantly underestimated the experimental resonance frequency, with errors of 27.14\% for the proposed model and 38.8\% for Vendik's model, while the mean squared errors were the highest (1.873\% and 2.253\%, respectively). Finally, Capacitor 3 showed the best agreement between theory and experiment. The mean squared errors were minimal (0.195
\% and 0.154\%), and the predicted resonance frequencies (575.7 MHz and 570.5 MHz) closely matched the experimental measurement (561.6 MHz).

\begin{table}[h!]
\caption{Mean Squared Error of |$Z_{11}$| Between Model Simulations and Physical Capacitors}
    \centering
    \renewcommand{\arraystretch}{1.5} % Aumenta o espaçamento entre as linhas
    \setlength{\tabcolsep}{5pt} % Ajusta o espaçamento entre colunas
    \small
    \begin{tabular}{c c c}
        \hline
\textbf{Capacitor Design} & \textbf{Proposed Model(\%)} &  \textbf{Vendik's Model(\%)}\\
        \hline
        1 & 1.060  & 1.323  \\
        2 & 1.873  & 2.253 \\
        3 & 0.195  & 0.154 \\

        \hline
    \end{tabular}
    
    \label{tab:erro}
\end{table}

\begin{table}[h!]
    \caption{Measured and Simulated Resonance Frequencies (MHz) for Physical and Simulated Capacitors}
    \centering
    \renewcommand{\arraystretch}{1.5} % Aumenta o espaçamento entre as linhas
    \setlength{\tabcolsep}{2.5pt} % Ajusta o espaçamento entre colunas
    \small
    \begin{tabular}{c c c c c c c c c c c c}
        \hline
\textbf{Capacitor Design}& \textbf{Measured} & \textbf{Proposed Model} &  \textbf{Vendik's Model}\\
        \hline
        1 & 352.4  & 407.1 & 427.4  \\
        2 & 537.1  & 391.3 & 339.3 \\
        3 & 561.6  & 575.7 & 570.5 \\
        \hline
    \end{tabular}

    \label{tab:freq}
\end{table}

%Analisando as Figs. \ref{fig:imagem_S11} e \ref{fig:imagem_Z11}, podemos perceber que o modelo RLC em série apresentado na seção \ref{sec:methods}, e em \cite{neu2003designing,winslow2000component} é o mais adequado para a faixa de operação proposto neste trabalho.

The comparative analysis demonstrates that both models show good agreement with experimental data for Capacitors 1 and 3, indicating similar capacitance modelling capabilities for these configurations. However, for Capacitor 2, the proposed model shows superior performance, with significantly reduced mean squared error compared to Vendik's model, which exhibits more pronounced discrepancies. These results demonstrate that while both methods are suitable for conventional geometries, the proposed model offers a significant advantage by incorporating additional physical effects that enhance overall modelling accuracy, particularly in configurations where Vendik's model exhibits limitations.

It is worth noting that the observed discrepancies, particularly for Capacitor 2, may be attributed to unmodelled experimental factors. The CNC fabrication process introduces dimensional variations, while the connection traces ($W$ = 1 mm) add unmodelled series resistance and parasitic inductance. For Capacitor 2 ($W$ = 4 mm), the connection traces represent $25\%$ of the total width. Furthermore, this model disagreement can be explained by geometric limitations, which warrant further investigation in future studies.

%% file: text/conclusion.tex
\section{Conclusion} 
\label{sec:conclusion}

This work proposed the development of a novel simplified planar capacitor model consisting of two identical rectangular plates, incorporating the effects of both the substrate and the medium between conductor traces on the total capacitance. The main objective was to establish a model capable of representing the behaviour of these components at frequencies below 1 GHz, while accounting for other factors that influence the impedance at high frequencies.

By doing so, we establish an important theoretical milestone by proposing a more comprehensive formulation that integrates three fundamental aspects: enhanced capacitance calculation (including substrate and inter-conductor medium effects), parasitic inductance modelling, and resistive loss consideration. This unified approach serves as a foundation for future research and practical applications in high-frequency circuit design, providing valuable tools for the design and analysis of planar capacitors and allowing the study of the phenomenon and its physical meaning directly from the modelled parameters.

Furthermore, the study demonstrated the importance of modelling both the dielectric substrate and the inter-conductor medium, with results showing significant improvements in the proposed model's accuracy compared to Vendik's model. Experimental validation revealed good agreement for most cases studied. Even for Capacitor 2, which showed the poorest agreement, the proposed model still outperformed Vendik's, confirming the robustness of the developed approach. These results highlighted the need for more detailed studies on geometric influences on model behaviour and the necessary adaptations in this regard.